\documentclass[pre,twocolumn,a4paper,showpacs]{revtex4-1}
\usepackage{amssymb}
\usepackage{wasysym}
\usepackage{graphicx}
\usepackage{epsfig}
\usepackage{subfigure}
\begin{document}
\title{Reputation blackboard systems }

\author{Jos\'e F. Fontanari}

\affiliation{Instituto de F\'{\i}sica de S\~ao Carlos,
  Universidade de S\~ao Paulo,
  Caixa Postal 369, 13560-970 S\~ao Carlos, S\~ao Paulo, Brazil}

\begin{abstract}
Blackboard systems are motivated by  the popular view of  task forces  as   brainstorming groups  in which specialists  write promising ideas   to solve a problem  in a central blackboard. Here we study a minimal model of  blackboard system designed  to solve  cryptarithmetic puzzles, where hints are posted anonymously  on a public display (standard blackboard)  or are posted together with information about the reputations of the agents that posted them (reputation blackboard).  We find that the reputation blackboard always outperforms the standard blackboard,  which, in turn, always outperforms the independent search.  The  asymptotic distribution of the computational cost of the search, which is proportional to the total number of agent updates required to find the solution  of the puzzle, is an exponential distribution for those three search heuristics. Only for the reputation blackboard  we  find a nontrivial dependence of the mean computational cost on the system size and, in that case, the optimal performance is achieved by a single agent working alone, indicating that, though the blackboard
organization can produce  impressive  performance gains  when compared with the independent search, it is not very supportive of cooperative work.
\end{abstract}
\maketitle
\section{Introduction}\label{sec:intro}

Understanding the conditions that improve the efficacy of  cooperative work  is a critical issue for the economy  of developed countries, given the central role played by task force problem-solving (e.g., drug design, robotics engineering, software development, etc.)  on producing technological innovations \cite{Page_07}. For instance, the study of  the impact of imposed communication patterns (i.e., who can communicate with whom) on group performance dates back to the 1950s \cite{Bavelas_50,Leavitt_51,Guetzkow_55} (see \cite{Mason_12,Reia_17} for more recent contributions) and  aims at offering  scientific guidance for matching  organizational structures 
 to  task complexity \cite{Mohr_82}. 

A classical team organization that allows relevant information (hints) to diffuse quickly among  members of the group is the so-called blackboard system \cite{Osborn_53}, which was introduced in the Artificial Intelligence  domain in the 1980s to tackle the uncertainties inherent to speech understanding \cite{Erman_80} and is now part of the AI problem-solving toolkit \cite{Englemore_88,Corkill_91}. In this organization, team members (or agents) simply read and write hints to a central blackboard that can be accessed by all members.

In this contribution we revisit and build on a seminal study of the efficacy of a  minimal model  of blackboard system to solve  cryptarithmetic problems \cite{Clearwater_91} (see also \cite{Clearwater_92}). That  study suggested that the blackboard organization could produce a superlinear speedup in the time to find the solution with respect to the number of group members.  We find that although  this organization indeed entails a considerable quantitative  gain in comparison with the situation where the  agents  solve the problem independently of each other using  a simple random trial-and-test search, it  does not produce any qualitative change in the performance measures. In particular, the  expected value of  shortest time $T_M$  to find the solution decreases linearly with the number of agents $M$  and the asymptotic distribution of $T_M$ is an exponential distribution as in the case of the independent search.  In addition, when the performance of the system is measured by the  computational cost $C \propto M T_M$, which essentially yields the total number of agent updates until the solution is found,  it becomes practically independent of the system size,  as in the  independent search again. 

Here we  propose a slightly different and, perhaps, more realistic  blackboard organization, the so-called reputation blackboard system, in which  the value of a hint (i.e., the probability of the hint being selected from the blackboard) is determined by the reputation of the agent that posted it. The reputation of an agent is a measure  of the quality of its partial solutions to the problem. We find that the  reputation blackboard   always outperforms the standard blackboard, and that the probability distribution of the computational costs are still well fitted by an exponential distribution. However, in contrast with the results for the independent search and for the standard blackboard, the mean computational cost exhibits a complex dependence on the system size $M$ and, somewhat surprising, the minimum cost is achieved for
$M=1$, i.e.,   individual work  is more effective than group work for this organization. 
Our findings suggests that the blackboard organization is not very supportive of cooperative work.

The rest of the paper is organized as follows. In Section \ref{sec:CP} we present  the cryptarithmetic problem used in our study  and explain how the solutions are encoded in integer strings. In that section we  also introduce the concept of hint and define  the cost function of a string, which is used to assign a reputation to the agents.  In Section \ref{sec:black}  we present the minimal model for 
the standard blackboard organization  \cite{Clearwater_91} and study its performance on solving the cryptarithmetic puzzle. Then in Section \ref{sec:rep} we  introduce the reputation blackboard organization  and compare its performance with that of the standard blackboard.
 Finally, Section \ref{sec:conc} is reserved to our concluding remarks.

\section{The cryptarithmetic puzzle}\label{sec:CP}

Cryptarithmetic problems such as 
\begin{equation}\label{DGR}
DONALD + GERALD = ROBERT 
\end{equation}
are constraint satisfaction problems in which the task is to find unique digit-to-letter assignments  so that the integer numbers represented by the words add up correctly \cite{Averbach_80}. In the cryptarithmetic problem  (\ref{DGR}), there are $10!$ different digit-to-letter assignments, of which only one is the solution to the problem, namely, $A=4, B=3, D=5, E=9, G=1, L=8, N=6, O=2, R=7, T=0$.  This type of cryptarithmetic problem, in which letters form meaningful words, are also termed alphametics \cite{Hunter_76} and were popularized in the 1930s by the {\it Sphinx}, a Belgian journal of recreational mathematics \cite{Averbach_80}. Of course, from the perspective of evaluating the performance of search heuristics the meaningfulness of the words is inconsequential. In this contribution we will focus  on the alphametic problem  (\ref{DGR}) because its state space is the largest possible (a cryptarithmetic puzzle must have at most $10$
different letters) and because it facilitates the independent replication of our findings.

For the alphametic problem   (\ref{DGR}) we encode a digit-to-letter assignment by the string 
${\mathbf i} = \left ( i_1,i_2, \ldots, i_{10} \right )$ where
$i_n = 0, \ldots, 9$ represent the 10 digits and the subscripts $n=1, \ldots, 10$ label the letters according to the convention 
\begin{eqnarray}\label{convention}
1 & \to &  A \nonumber \\
2 & \to & B \nonumber \\
3 &\to  & D \nonumber \\
4 &\to  & E \nonumber \\
5 & \to & G \nonumber \\
6 &\to & L \nonumber \\
7 &\to &N \nonumber \\
8 &\to &O \nonumber \\
9 &\to &R \nonumber \\
10 &\to & T .
\end{eqnarray}
%
 For example, the
string $\left ( 0,2,9,4,8,1,7,6,3,5 \right)$ corresponds to the digit-to-letter assignment $A=0, B=2, D=9, E=4, G=8, L=1, N=7, O=6, R=3, T=5$.
Some search heuristics require that we assign a cost to each string, which then is viewed as a measure of the goodness of the partial answer represented by the string. A natural choice for the cost function is \cite{Abbasian_09,Fontanari_14}
\begin{equation}\label{cost}
c \left ( {\mathbf i} \right ) = \left | R - \left ( F + S \right ) \right |
\end{equation}
where $R$ is the result of the operation ($ROBERT$), $F$ is the first operand ($DONALD$) and $S$ is the second operand ($GERALD$).
For the string $\left ( 0,2,9,4,8,1,7,6,3,5 \right)$ we have $R = 362435$, $F= 967019$ and $S = 843019$ so that the cost associated  is $c = 1447603$. The cost value defined in eq. (\ref{cost})
applies to all strings except those for which $i_3 = 0$ corresponding to the assignment $D=0$, $i_5 = 0$ 
corresponding to the assignment $G=0$ and $i_9 = 0$ corresponding to the assignment $R=0$. Those are  invalid strings because
they violate the rule of the cryptarithmetic puzzles that an integer number should not have the digit $0$ in its leftmost position. Hence for
those strings we assign an arbitrary large cost value, namely, $c = 10^8$, so that now they become valid strings but  they have the
highest  cost among all strings. If the cost of a string is $c=0$ then the digit-to-letter assignment coded
by that string is the  solution to the cryptarithmetic problem.

An advantage of traditional blackboard systems is that they do not  need the introduction of a  cost function to weight the quality of
the strings. Rather, those systems use  hints that the agents read and write to a blackboard that is accessed by all agents \cite{Englemore_88}. In the context of  cryptarithmetic puzzles, hints are letter-digit assignments that add up correctly modulo $10$ for at least one column \cite{Clearwater_91,Clearwater_92}. For example, considering the third column (from left to right) of the problem  (\ref{DGR}) we have the hints $\left ( N=3, R=2, B=5 \right )$,  $\left ( N=7, R=8, B=5 \right )$, etc. 
Although this is the  definition of hints used in Refs.\  \cite{Clearwater_91,Clearwater_92}, it is actually not very useful to solve the puzzle 
 (\ref{DGR}) since the correct solution uses only 2 of those hints, namely, $\left ( N=6, R=7, B=3 \right )$ and $\left ( D=5, D=5, T=0 \right )$, whereas the other 4 columns  only add up correctly if one considers the $1$ that is carried from the adjacent column, e.g., $\left ( L=8, L=8, R=7 \right )$, $\left ( A=4, A=4, E=9 \right )$, etc. This observation suggests that we extend the list of hints to include also the cases that the two letter-digit assignments plus the carry 1 add up correctly modulo $10$ for each column, except for the leftmost one ($D+D=T$), of course. In this scenario, there are $351$ distinct hints and $6$ of them correspond to the  solution of the puzzle   (\ref{DGR}). Here we will consider 
only the extended list of hints.

As we will see in the next section, the hints are discovered by the agents during their random exploration of the state space and displayed in
the central  blackboard for use by the other agents.

%
\begin{figure}
\centering
\includegraphics[width=0.48\textwidth]{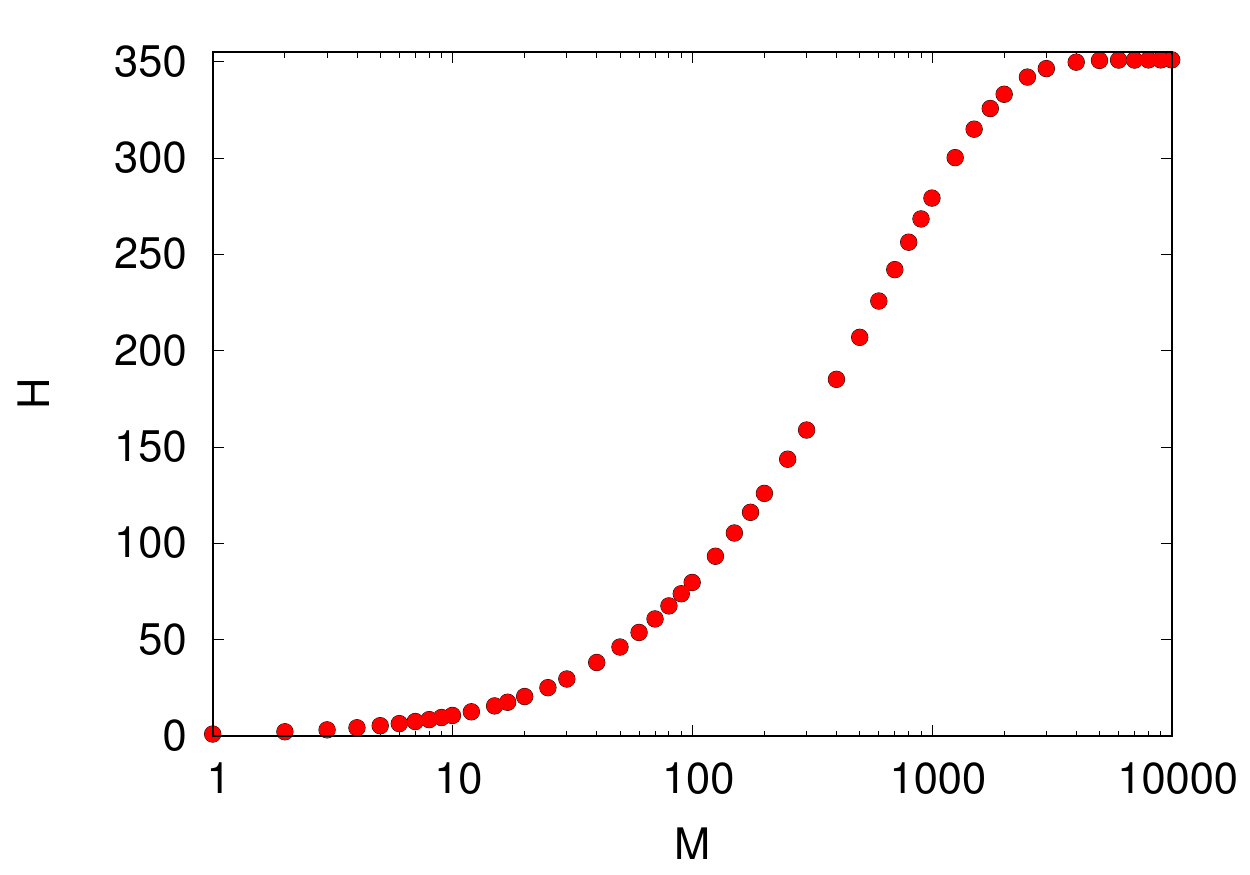}
\caption{(Color online) Mean number of distinct hints $H$ displayed on the blackboard at the beginning of the search as function of the system size $M$.  The mean was estimate by generating $10^5$  systems of size $M$. The maximum number of distinct hints for the cryptarithmetic problem  (\ref{DGR}) is $351$. }
\label{fig:1}
\end{figure}
%

\section{The standard  blackboard system}\label{sec:black}

We consider a  system  composed of $M$  agents and a central blackboard where the agents can read and write hints. However, hints cannot be erased from the  blackboard.  Each agent is represented by a string of $10$ different digits representing
a particular digit-to-letter assignment for the puzzle  (\ref{DGR}) and so  henceforth  we will use the terms agent and string interchangeably. At time $t=1$ all agents are initialized as random strings selected with equal probability from the pool of the $10!$ valid digit strings. The agents check for all possible hints from their strings and post them to the blackboard, unless they are already displayed on the board. Figure \ref{fig:1} shows the mean number of hints $H$ displayed  on the board at time $t=1$ for different system sizes $M$. For instance, for $M=1$ we have $H \approx 1.1$, for $M=100$, $H \approx 80$  and for $M > 5000$ 
the board will almost certainly display all $351$ hints already at the outset of the search.
 
Once the initial states of the agents and of the blackboard are set up, the search procedure develops as follows. It begins with  a randomly chosen agent -- the target agent -- picking a  hint at random from the  blackboard. In the case that there are no hints (i.e, the blackboard is empty), or that the target agent is already using the chosen hint,  the  agent selects randomly  a letter-to-digit assignment from the pool of  valid digit strings. The incorporation of a hint into the digit string of the target agent involves the relocation of at most six digits of that string.
For example, consider the assimilation of the hint  $\left ( N=1, R=4, B=5 \right )$ into the target string $\left ( 0,2,9,4,8,1,7,6,3,5 \right)$. First,   the assignment $N=1$ is assimilated, yielding $\left ( 0,2,9,4,8,7,1,6,3,5 \right)$, then $R=4$, yielding $\left ( 0,2,9,3,8,7,1,6,4,5 \right)$ and finally $B=5$ resulting in the string $\left ( 0,5,9,3,8,7,1,6,4,2 \right)$ which contains the desired hint. In both events -- incorporation of a hint into the target string or replacement of the target string by a random string -- the agent checks for all possible hints from its new string and posts them to the blackboard.

After the target agent  is updated, we increment the time $t$ by the quantity $\Delta t = 1/M$.  Then another target agent is selected at random and the procedure described before is repeated. Note that during the increment from $t$ to $t+1$ exactly  $M$, not necessarily distinct, agents are updated. Let us denote by $t_i^* = 1,2, \ldots$ the length of time that agent $i$  takes  to find the solution of the cryptarithmetic problem.
Although Refs.\ \cite{Clearwater_91,Clearwater_92} have focused on the
distribution of $t_i^*$,  we think that a more suitable measure of the efficiency of the blackboard system is  
\begin{equation}\label{TM}
T_M = \min \left ( t_1^*, \ldots, t_M^* \right ),
\end{equation}
 which  is interpreted as the first  time that the solution of the puzzle is  found by one of the agents. 

In the case the agents explore the state space independently of each other, the $t_i^*$s,  with 
$i=1, \ldots,M$, are  identically distributed  independent random variables distributed by the geometric distribution
\begin{equation}
f \left ( t_i^* \right ) =  p \left ( 1- p \right)^{t_i^*-1},
\end{equation}
where $p= 1/10!$ is the success probability. (We recall that the puzzle (\ref{DGR}) has a unique solution.) The probability distribution of the minimum time $T_M$ is also a geometric distribution
\cite{Feller_68,Fontanari_15} with success probability $1 - \left ( 1 - p \right)^M$,
\begin{eqnarray}\label{PTM}
\tilde{P} \left ( T_M \right ) & = & \left [ 1 - \left ( 1 - p \right)^M \right ] \left ( 1 - p \right)^{M \left ( T_M - 1 \right )} \label{PTM0} \\
& \approx & M p \exp \left ( - Mp T_M \right )  \label{PTM},
\end{eqnarray}
where in the last step we have assumed that the system size is much small\-er than the size of the state space, i.e., $Mp \ll 1$.  

A useful measure of the performance of a  problem-solving system is the  computational cost of the search defined as
\begin{equation}\label{C}
C  =  M p  T_M ,
\end{equation}
which yields the total number of agent updates necessary to find the solution scaled by the effective size of the state space. Since $ P \left ( C  \right ) = \tilde{P} \left ( T_M \right )/Mp $ we have  $ P \left ( C  \right ) = \exp \left ( - C \right )$  for 
the independent search with $Mp \ll 1$. The advantage of using $C$  is that the dependence of its  mean on the number of agents  can be used to distinguish between truly cooperative  from non-cooperative systems, for which $\langle C \rangle$ is approximately  constant. 

%
\begin{figure}
\centering
\includegraphics[width=0.48\textwidth]{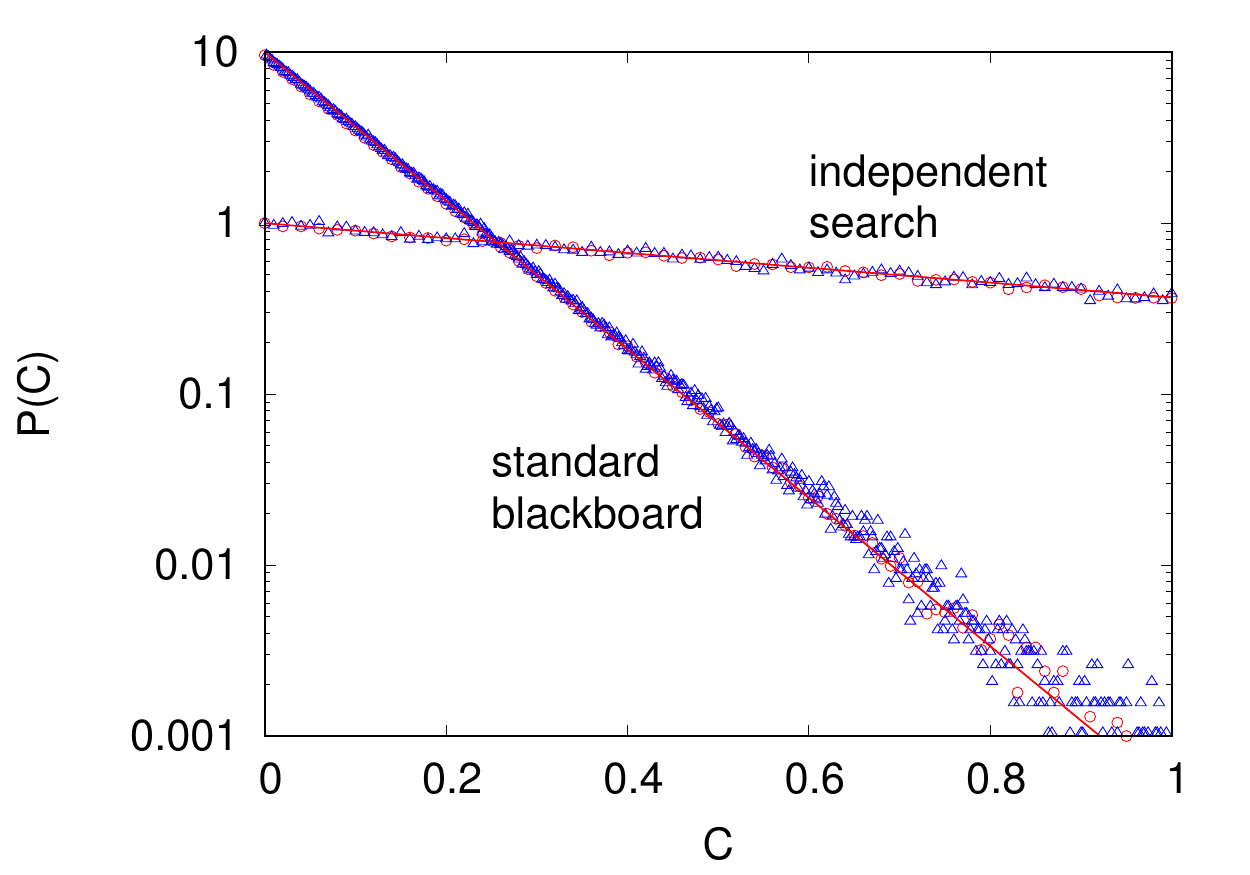}
\caption{(Color online) Probability distribution of the computational cost $C = M T_M/10!$ to find the solution of the puzzle (\ref{DGR}) for  the independent search and  the standard blackboard system as indicated. The  system sizes are  $M=1$ ($\triangle$) and $M=100$ ($\circ$).
The curve fitting the data of the independent search is $P \left ( C  \right ) = \exp \left ( - C \right )$, whereas the data of the
blackboard system is fitted by $P \left ( C  \right ) = 10 \exp \left ( - 10 C \right )$.   }
\label{fig:2}
\end{figure}
%

Figure \ref{fig:2} shows the  distribution of probability $ P \left ( C \right ) $ of  the computational cost  for the independent search and for the standard blackboard system in the realistic situation that $Mp \ll 1$.  As expected, the results for the independent search  are  independent of the system size $M$.
We find the same (qualitative) results for the standard blackboard system: although this organization reduces the time to solve the puzzle by a factor of 10,  the mean computational cost  $\langle C \rangle  \approx 0.1 $ is practically unaffected by the value of $M$, provided it is not  too large (see Fig.\ \ref{fig:5}),  as in the case of the  independent search. This is not a surprise since the results shown in Fig.\ \ref{fig:1} indicate that  the blackboard will display all hints after a very short time. In particular,   in the case of a single agent ($M=1$), it takes on the average only $t/10! = 0.0035$ updates to fill out the blackboard. Hence 
the cooperation (i.e., writing on the board) ceases at the very beginning of the search and from then on the $M$ agents will explore the state space and  the changeless blackboard independently of each other. We estimated that the replacement of the target string by a random string occurs with probability $0.0068$, so most of the time the target agents are picking hints from the blackboard.  

Our results indicate that although the standard blackboard organization can produce a tenfold  speedup on the mean time to find the solution of the cryptarithmetic puzzle, it does not   change the nature of the search, which maintains all characteristics of the
independent search, contrary to the suggestion of  Refs.\ \cite{Clearwater_91,Clearwater_92} that the exponential distribution of eq.\ (\ref{PTM}) is replaced by a lognormal distribution for the blackboard organization. The main point, however, is that such   organization is not really a cooperative problem-solving system, since once the blackboard is filled out, which happens in a very short time, the agents will pick  hints on the board and  explore the state space independently of each other.

\section{The reputation  blackboard system}\label{sec:rep}

A straightforward  way to guarantee that the agents keep updating the blackboard with useful information even after all hints are already on display is to associate an indicator of quality to each hint. This indicator is the reputation of the target agent who wrote the hint, and is measured by the  reciprocal of the cost function (\ref{cost}). (The fact that the cost is zero for the solution of the puzzle is not a problem because the search is halted  when the solution is found  as we are interested in the statistics of $T_M$ only.) 
Rather than picking hints from the board at random  with equal probability as done in the standard blackboard scenario of the previous section,
now the probability of selecting a hint is proportional to the reputation of the hint, so that hints posted by high-reputation agents are  more influential than those posted by low-reputation agents, a situation  that resembles the Ortega hypothesis about the scientists contributions to scientific progress \cite{Cole_72}.

The search procedure for the reputation blackboard  is almost the same as for the standard blackboard, except that the  hints are displayed on the board together with the reputation of the agents who last wrote them, and that the probability of selecting a hint is directly proportional to the reputation of the agent that posted it, as already mentioned.
In particular, we assume that the target agent overwrites the hints on the blackboard so they will be displayed with its reputation until  another
agent overwrites them again. Of course, after the short time necessary to fill out the board with the 351 hints,  only the reputation labels  will change in the rest of the search.  

%
\begin{figure}
\centering
\includegraphics[width=0.48\textwidth]{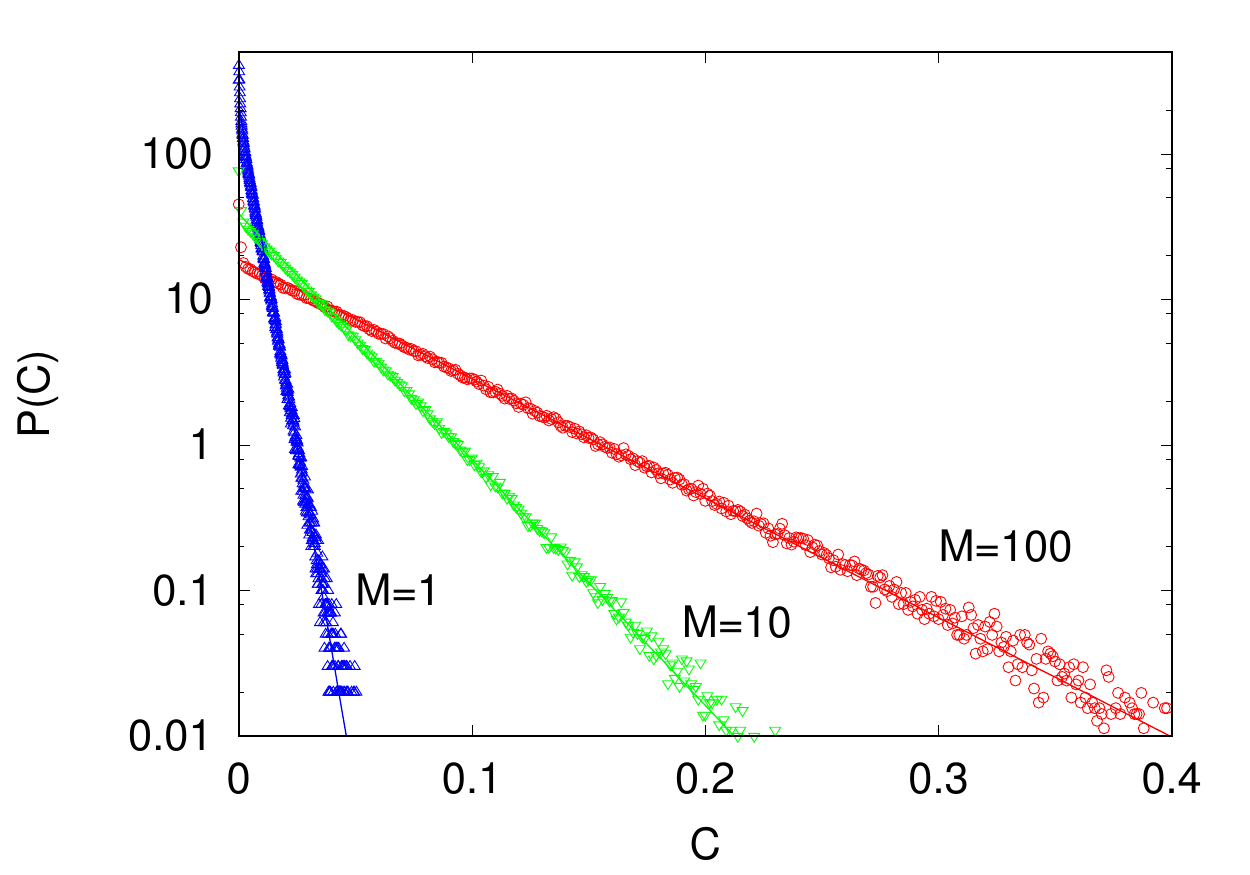}
\caption{(Color online) Probability distribution of the computational cost $C = M T_M/10!$  to find the solution of the puzzle  (\ref{DGR}) for  reputation blackboards of sizes  $M=1$ ($\triangle$), $M=10$ ($\triangledown$) and $M=100$ ($\circ$) as indicated. Here the solid lines are the fittings using the exponential distribution $P \left( C \right ) =\exp \left ( - C/\lambda_M \right )/\lambda_M$ with $\lambda_1 = 0.0046$,
$\lambda_{10} = 0.026$ and $\lambda_{100} = 0.053$,
}
\label{fig:3}
\end{figure}
%

\begin{figure}
\centering
\includegraphics[width=0.48\textwidth]{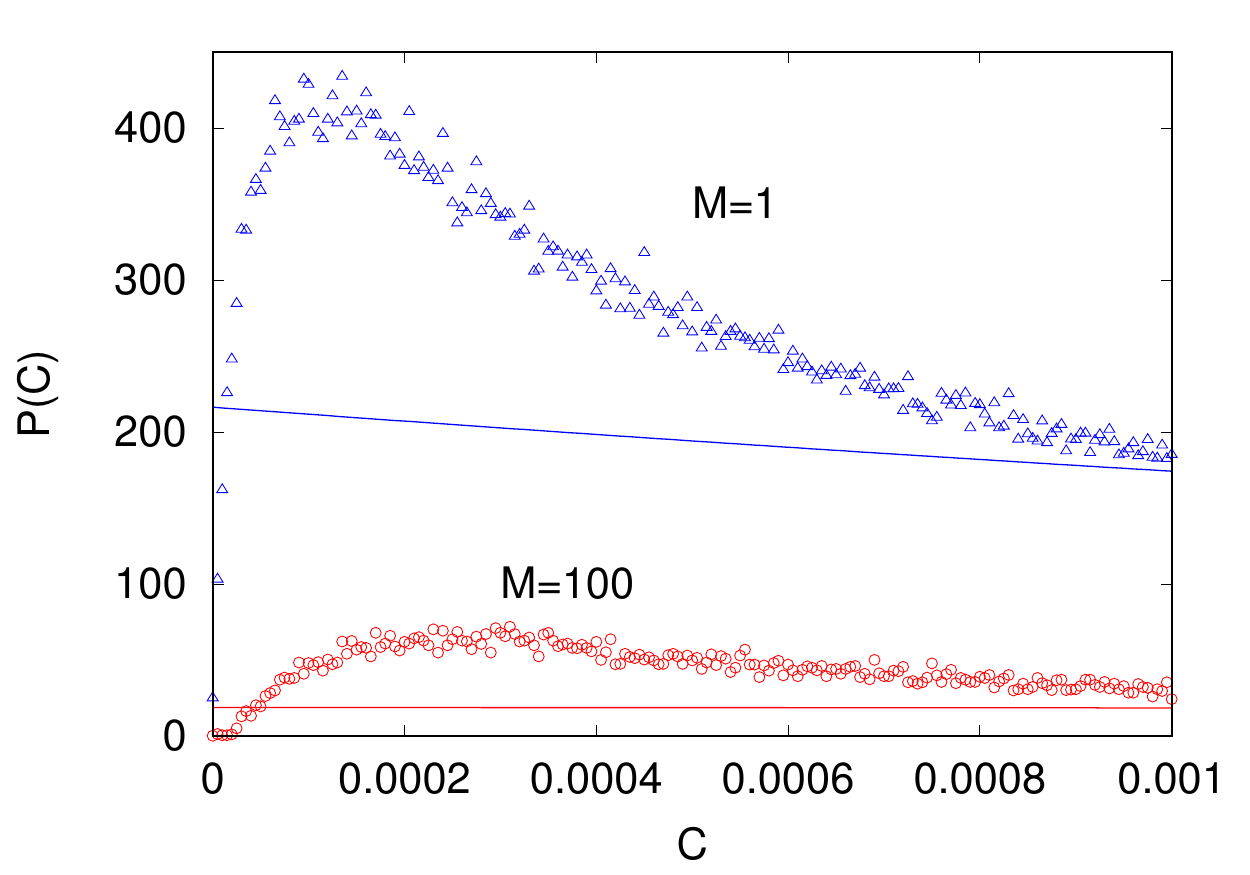}
\caption{(Color online) Close-up view of the low cost region of Fig.\ \ref{fig:3}  for  system sizes   $M=1$ ($\triangle$)  and $M=100$ ($\circ$) as indicated. 
}
\label{fig:4}
\end{figure}

Figure \ref{fig:3} shows the distribution of the computational cost for  reputation blackboard systems of  different sizes. In contrast to the results for the independent search and for the standard blackboard (see Fig.\ \ref{fig:2}), these results show   a marked dependence on the system size $M$ that points to the collective nature of the search procedure. The  distribution $P \left ( C \right )$ in the regime of not too small costs  is well fitted by an  exponential, which is a robust marker of the essentially random trial-and-error character of the search. However, for small computational costs we find significant deviations from the exponential fitting as shown in Fig.\ \ref{fig:4}. This is probably due to the fact that the blackboard does not yet  exhibit all the 351 hints  in the short time region depicted in the figure. We find a similar  but much less pronounced effect in the case of the standard blackboard as well. We note that the higher odds of finding the solution for low cost searches when compared with  the predictions of the exponential fitting has only a minor effect on the mean computational cost, as we will see next.

%

\begin{figure}
\centering
\includegraphics[width=0.48\textwidth]{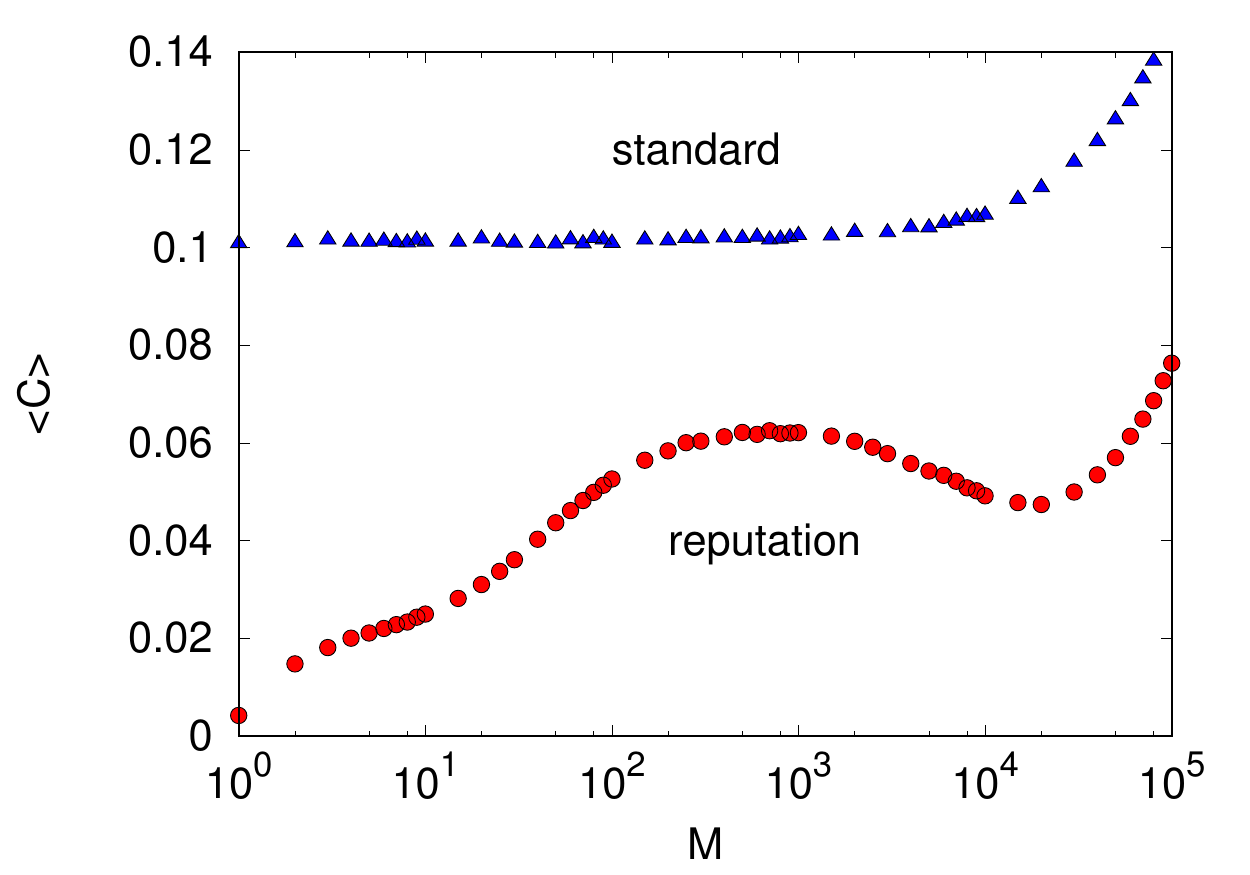}
\caption{(Color online) Mean computational cost $\langle C \rangle $  to find the solution of the puzzle $DONALD + GERALD = ROBERT$ as function of the system size $M$ for the standard ($\blacktriangle$) and the reputation ($\CIRCLE$) blackboards as indicated. The error bars are smaller than the symbol sizes.  }
\label{fig:5}
\end{figure}

A more revealing comparison between the standard and reputation blackboards is offered in Fig.\ \ref{fig:5} that shows the mean computational cost as function of the system size. Whereas the mean cost of  the standard blackboard system remains constant within a vast range of system sizes and begins to increase noticeably only for $M > 10^4$ due to duplication of work (i.e., different agents searching the same region of the state space), the mean cost of the reputation blackboard system exhibits a somewhat complex behavior. In particular, the increase of $\langle C \rangle $  for small $M$ is probably the effect of  hints posted by high-reputation agents trapped in local 
minima of the cost function (\ref{cost}). When the population is large enough this effect is attenuated by the presence of hints posted by agents close to the global minimum (solution), but then the duplication of work  ends up increasing the computational cost again. These phenomena were observed in the study of imitative learning, for which the local optima play a major role  \cite{Fontanari_15,Fontanari_16}. The novelty here is that the best performance is achieved by  a single individual  ($M_{opt} =1$), whereas in the imitative learning case the
optimal performance  is achieved by a group size that is on the order of the logarithm of the size of the state space (i.e, $M_{opt} \approx \ln 10! \approx 15$) \cite{Fontanari_14}. 

The mean cost for $M=1$ is $\langle C \rangle  \approx 0.0042$, which shows that the reputation blackboard  is about 20 times more efficient than the standard blackboard and 200 times more efficient than
the independent search.  As expected,  this mean cost is slightly lower than the mean of the exponential distribution $\lambda_1$ given in the caption of Fig.\ \ref{fig:3}. In this case, the blackboard functions as  the  agent's memory,  which is used to track its past successes and failures.

 %
%
\begin{figure}
\centering
\includegraphics[width=0.48\textwidth]{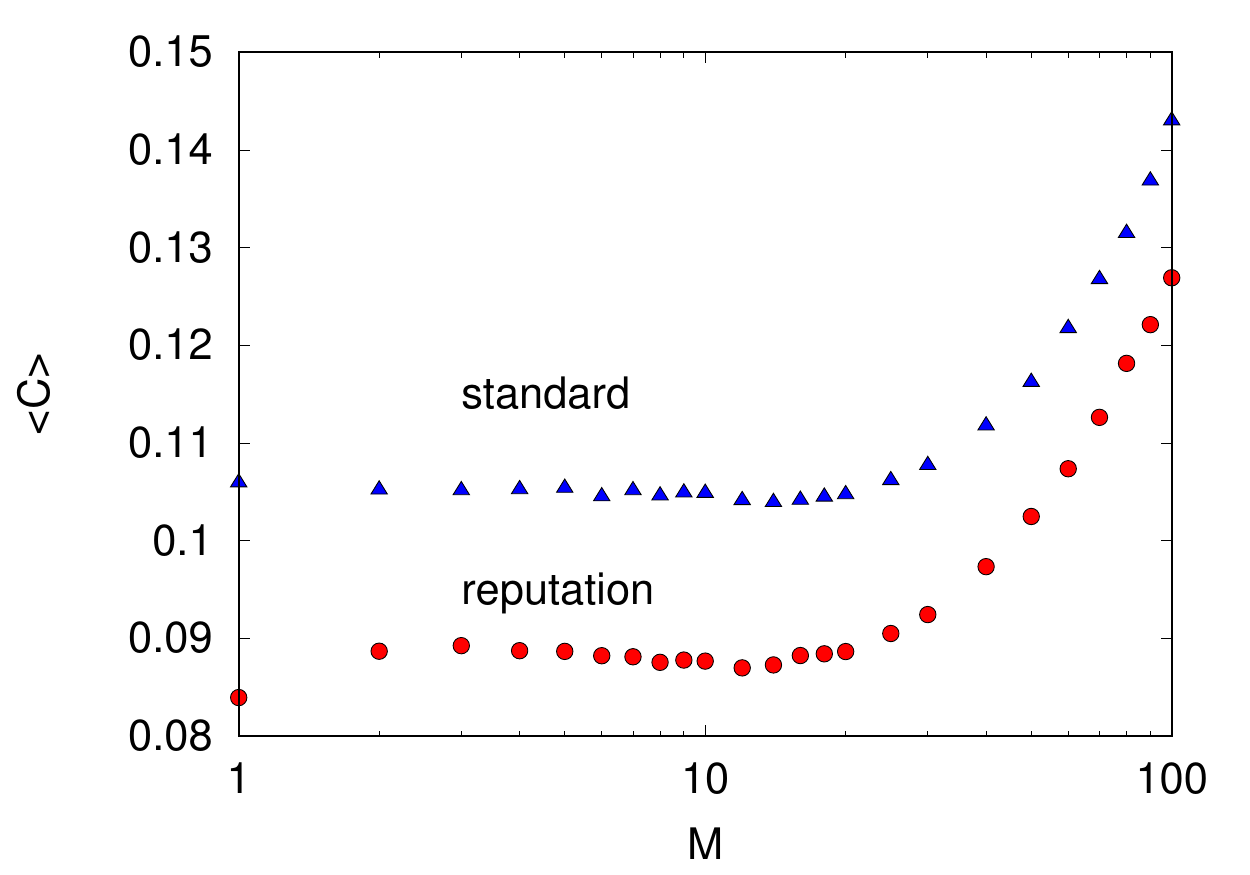}
\caption{(Color online) Mean computational cost $\langle C \rangle $  to find the solution of the puzzle $WOW + HOT = TEA$ as function of the system size $M$ for the standard ($\blacktriangle$) and the reputation ($\CIRCLE$) blackboards as indicated. The error bars are smaller than the symbol sizes.  }
\label{fig:6}
\end{figure}

Finally, we note that,  qualitatively speaking, our conclusions hold true for other cryptarithmetic puzzles as well, provided the computational cost is defined correctly. For instance, consider the popular puzzle  
\begin{equation}\label{WHT}
WOW + HOT = TEA,
\end{equation}
whose state space is comprised of the $10!/4! = 151200$  sequences of 6 distinct digits. This problem has 66 different solutions (we have generated and tested all possible sequences) so that the probability of finding the solution by picking a single sequence at random  is $p = 66/151200$. Inserting  this value of $p$ in eq.\ (\ref{C})  
allows us to compare the performances of the search heuristics on different puzzles since the effective sizes of their state spaces (i.e., $1/p$) are accounted for in the definition of the computational cost.  Figure \ref{fig:6}, which shows the mean computational cost for puzzle (\ref{WHT}),
confirms the general validity  of our conclusions. We recall that for the independent search one has 
$\langle C \rangle = Mp/ \left [ 1 -  \left ( 1-  p \right )^M \right ]$  so that $\langle C \rangle \approx 1$ for $Mp \ll 1$. Interestingly,  if we  estimate the difficulty of a puzzle  by the computational cost to find its
solutions, then,  from the perspective of the blackboard systems,  puzzle (\ref{WHT})  is  more  difficult than  puzzle (\ref{DGR}).

\section{Discussion}\label{sec:conc}

The appeal of the blackboard organization probably owns to the common view of  working groups as  teams of specialists  exchanging ideas on possible approaches to solve a problem and writing the promising lines of investigation in a blackboard \cite{Corkill_91}. A  minimal model of blackboard  systems \cite{Clearwater_91,Clearwater_92}, which we refer to as the standard blackboard organization, assumes that the agents post hints anonymously on the blackboard, and that  those hints have equal probability of selection or, equivalently, have equal value.  The performance of the system is measured by the computational cost of the search, which is  proportional to the  total number of agent updates required to find the  solution of the problem posed to the group (see eqs.\ (\ref{TM}) and (\ref{C})).

Our findings  show  that the  standard blackboard organization
produces a tenfold decrease on the mean computational cost to find the solution of the cryptarithmetic puzzle (\ref{DGR}) when compared with the case that the agents work independently of each other (see Fig.\ \ref{fig:2}). However, this cost is practically unaffected by changes on the number of agents $M$ for not too large $M$, as in the case of the independent search. This result reveals that  this organization is not very effective on promoting cooperative work. In fact, once all hints are displayed on the blackboard, which happens in a very short time span compared with the total length of the search, the agents  carry out the search independently of each other.     

In this contribution we propose a perhaps more realistic model of blackboard organization, referred to as the reputation blackboard, in which the value of a hint (i.e., the probability of the hint being selected from the blackboard) is associated to the reputation of the agent that posted it. The reputation of, say, agent $i$ is defined as the reciprocal of the cost function $c \left (i \right )$ given in eq.\ (\ref{cost}). We find that the computational cost of the reputation blackboard is always lower than the cost of the standard blackboard, and it exhibits a nontrivial dependence on the number of agents in the system (see Fig.\ \ref{fig:5}), which reveals the cooperative nature of the search. However, for both blackboard organizations
the probability distribution of the computational cost is an exponential distribution as in the case of the independent search. This observation is at variance with earlier findings \cite{Clearwater_91,Clearwater_92}, which predicted a lognormal distribution for the standard blackboard organization.

A most unexpected  and, perhaps, instructive outcome of our  analysis of the reputation blackboard is that the best performance is achieved for a single agent (i.e., $M=1$), which suggests that individual work is more efficient than group work in that case (see  Figs.\ \ref{fig:3} and \ref{fig:5}). The reason is that in our blackboard scenario all agents have the potential to solve the problem by themselves and in that case
the optimal strategy is to employ a single agent that uses the blackboard as a notebook  to keep track of its past failures and successes and decide its future actions, thus carrying out a sort of individual brainstorming. 

In a real-world scenario, it is likely that none of the team members are able to solve the problem working alone, otherwise hiring a single specialist would obviously be the cheapest way to carry out the task, as we have shown here. It is easy to modify the minimal model of blackboard organization to guarantee that the agents cannot solve the problem by themselves (e.g., by allowing the agents to search only  limited regions of the state space) and  so to make cooperation mandatory for solving the problem.  But then we would not be able to make a fair comparison between the performances of the  cooperative system and the independent search, which provides a valuable quantitative appraisal of  the efficiency of the  blackboard system. Nevertheless, it would be instructive to find out what ingredients one should add to our reputation blackboard scenario in order to make group work more efficient than individual work. 

From a more general perspective, we note that
finding a cooperative organization that produces  a superlinear speedup in the time to find the solution with respect to the number of agents remains a challenging issue for distributed cooperative problem-solving  systems  \cite{Huberman_90}.

\acknowledgments
This research  was  supported in part by grant
2015/21689-2, Funda\c{c}\~ao de Amparo \`a Pesquisa do Estado de S\~ao Paulo 
(FAPESP) and by grant 303979/2013-5, Conselho Nacional de Desenvolvimento 
Cient\'{\i}\-fi\-co e Tecnol\'ogico (CNPq).


\begin{thebibliography}{99}

\bibitem{Page_07}
S.E. Page, 
\textit{The Difference: How the Power of Diversity Creates Better Groups, Firms, Schools, and Societies} 
(Princeton University Press, Princeton, NJ, 2007)

\bibitem{Bavelas_50}
A. Bavelas, 
J. Acoustical Soc. Amer.  \textbf{22},   725--730 (1950)

\bibitem{Leavitt_51}
H.J. Leavitt, 
J. Abnorm. Soc. Psych.  \textbf{46} 38--50 (1951)  

\bibitem{Guetzkow_55}
H. Guetzkow, H.A. Simon,  
Manage. Sci. \textbf{1},    233--250 (1955)

\bibitem{Mason_12}
W. Mason, D.J. Watts,  
 Proc. Natl. Acad. Sci.  \textbf{109} 764--769  (2012) 
 
 \bibitem{Reia_17}
 S.M. Reia, S. Herrmann, J.F. Fontanari,
 Phys. Rev. E  \textbf{95}, 022305  (2017)
 
 \bibitem{Mohr_82}
 L.B. Mohr,  
 \textit{Explaining Organizational Behavior}
 (Jossey-Bass Publishers, San Francisco, 1982)
 
 \bibitem{Osborn_53}
 A.F. Osborn,
 \textit{Applied Imagination: Principles and Procedures of Creative Problem Solving}
 (Charles Scribner's Sons, New York, 1953)

\bibitem{Erman_80}
L.D. Erman, F. Hayes-Roth, V.R. Lesser, D.R. Reddy,
ACM Comput. Surv. \textbf{12}, 213--253 (1980)

\bibitem{Englemore_88}
 R. Englemore, T. Morgan,
\textit{Blackboard Systems}
 (Addison-Wesley, New York, 1988)
 
\bibitem{Corkill_91}
D.D. Corkill,  
AI Expert \textbf{6}, 40--47 (1991)

\bibitem{Clearwater_91}
S.H. Clearwater,  B.A. Huberman,  T. Hogg, 
Science \textbf{254}, 1181--1183 (1991)

\bibitem{Clearwater_92}
 S.H. Clearwater, T.  Hogg,  B.A. Huberman,  
in {\it Computation: The Micro and the Macro View}, edited by B.A. Huberman 
(World Scientific, Singapore, 1992),  pp. 33--70

\bibitem{Averbach_80}
  B. Averbach, O. Chein,
\textit{Problem Solving Through Recreational Mathematics}
 (Freeman, San Francisco, 1980)

\bibitem{Hunter_76}
 J.A. Hunter,
\textit{Mathematical Brain Teasers}
(Dover, New York, 1976)

\bibitem{Abbasian_09}
   R. Abbasian, M. Mazloom,
in  {\it  Proceedings of the Second International Conference on Computer and Electrical Engineering} 
(IEEE Computer Society, Washington, DC,  2009) pp. 308--312

\bibitem{Fontanari_14}
J.F. Fontanari, 
PLoS ONE \textbf{9}, e110517 (2014)

\bibitem{Feller_68}
W. Feller,
\textit{An Introduction to Probability Theory and Its
Applications}
Vol.  1  (Wiley,  New  York,  Third  Edition, 1968) p. 220

\bibitem{Fontanari_15}
J.F. Fontanari, 
Eur. Phys. J. B  \textbf{88},  251 (2015)

\bibitem{Cole_72}
J.R. Cole, S. Cole,
Science \textbf{178}, 368--374 (1972)

\bibitem{Fontanari_16}
J.F. Fontanari, 
EPL  \textbf{113}, 28009 (2016)


\bibitem{Huberman_90}
B.A. Huberman, 
Physica D \textbf{42}, 38--47 (1990)


\end{thebibliography}
\end{document}